\documentclass{IEEEtran}
\usepackage{cite}
\usepackage{amsmath,amssymb,amsfonts}
\usepackage{algorithmic}
\usepackage{algorithm}
\usepackage{algorithmic}
\usepackage{graphicx}
\usepackage{textcomp}
\usepackage{epstopdf}
\def\BibTeX{{\rm B\kern-.05em{\sc i\kern-.025em b}\kern-.08em
    T\kern-.1667em\lower.7ex\hbox{E}\kern-.125emX}}
\begin{document}
\title{Auto-Optimization with Active Learning in Uncertain Environment: A Predictive Control Approach}

\author{Yuan~Tan, Jun Yang,~\IEEEmembership{Fellow,~IEEE,} Zhongguo~Li,  Wen-Hua Chen, \IEEEmembership{Fellow,~IEEE}, Shihua Li, \IEEEmembership{Fellow,~IEEE}
\thanks{(Corresponding author: Jun Yang)}
\thanks{Yuan~Tan is with with the School of Electrical Engineering and Automation, Hubei Normal University, Huangshi, China  (e-mail: tanyuan@hbnu.edu.cn)}
\thanks{Jun Yang is with the Department of Aeronautical and Automotive Engineering, Loughborough University, Loughborough, England (e-mail: j.yang3@lboro.ac.uk)}
\thanks{Zhongguo~Li is with the Department of Electrical and Electronic Engineering, The University of Manchester, Manchester,  United Kingdom (e-mail: zhongguo.li@manchester.ac.uk)}
\thanks{Wen-Hua~Chen is with the Department of Aeronautical and Aviation Engineering, Hong Kong Polytechnic University, QR829, P. R. China. (e-mail:
wenhua.chen@polyu.edu.hk)}
\thanks{Shihua Li is with the School of Automation, Southeast University,  Nanjing, China (e-mail:lsh@seu.edu.cn)}
}

\markboth{ }%
{Shell \MakeLowercase{\textit{et al.}}: Bare Demo of IEEEtran.cls for IEEE Journals}
\maketitle

\begin{abstract}
This paper presents an auto-optimal model predictive control (MPC) framework enhanced with active learning, designed to autonomously track optimal operational
 conditions in an unknown environment,where the conditions may dynamically adjust to environmental changes.
First, an exploitation-oriented MPC (EO-MPC) is proposed, integrating real-time sampling data with robust set-based parameter estimation techniques to address the critical challenge of parameter identification. By introducing virtual excitation signals into the terminal constraint and establishing a validation mechanism for persistent excitation condition, the EO-MPC effectively resolves the issue of insufficient persistent excitation in parameter identification. Building upon this foundation, an active learning MPC (AL-MPC) approach is developed to  integrate both available and virtual future data to resolve the fundamental conflict between
 tracking an unknown optimal operational condition and parameter identification. The recursive feasibility and convergence of the proposed methods are rigorously established, and numerous examples substantiate the reliability and effectiveness of the approach in practical applications.
\end{abstract}
  \newtheorem{theorem}{Theorem}
\newtheorem{lemma}{Lemma}                            
   \newtheorem{assum}{Assumption}                            
\newtheorem{remark}{Remark}
\begin{IEEEkeywords}
Auto-Optimal Model Predictive Control, Active Learning, Exploration and Exploitation, Robust Set-based Parameter Estimation, Persistent Excitation.
\end{IEEEkeywords}

\section{Introduction}
\subsection{Motivation}
Set-point tracking model predictive control (MPC) has made significant progress, particularly in applications such as industrial process control, autonomous driving \cite{farag2020complex}, and robotic navigation \cite{hirose2019deep}. Traditional set-point tracking MPC methods primarily focus on optimizing control inputs to ensure that the system state accurately tracks the given reference signal while satisfying system constraints \cite{T11,simon2014reference}.

However, in many practical applications, set-points depend on unknown or variable environmental parameters, making it challenging to specify the set-points in advance.
Achieving optimal system performance is critical for maximizing efficiency, productivity, or profitability, yet this becomes increasingly difficult in unknown environments due to disturbances and unknown parameters. For example, heating control systems in buildings must adjust to fluctuating occupancy levels and external weather conditions to balance comfort and energy efficiency \cite{T1}. Wind turbines, meanwhile, adjust blade angles and rotation speeds in response to changing wind conditions to optimize power output \cite{T2}. In automotive systems, anti-lock braking systems (ABS) dynamically regulate brake pressure to maximize tractive forces between road surface and tyres, under different road conditions \cite{T3}. Similarly, in photovoltaic systems, maximum power point tracking (MPPT) adapts continuously to varying conditions to deliver the highest possible power output despite changes in temperature and solar irradiance \cite{42,43}. These examples underscore the necessity of control systems that can respond in real-time to dynamic changes in their operating environments, ensuring optimal performance \cite{T8,T27,T18}.

In light of these challenges, we propose an auto-optimal MPC framework with active learning, which addresses the need for real-time adaptation in uncertain environments. By incorporating active learning techniques, the framework can continuously refine its understanding of the environment, enabling the system to adjust its control strategy to achieve optimal operational conditions, even when faced with unknown or changing conditions.

\subsection{Related Work}
Auto-optimal MPC aims to keep the system operating at an optimal set-point, even in the face of unknown or constantly changing environmental parameters \cite{LiZ}. As the optimal operating conditions may be unknown and subject to change during the process, a control system must learn and adapt by interacting with its environment in real-time. Through limited but insightful interactions, the system gathers information and updates its strategies to track estimated set-points or references based on its growing knowledge of the environment.

Auto-optimal MPC involves two inherently conflicting objectives: parameter identification and optimality tracking. While accurately identifying unknown parameters is crucial for enhancing system performance, this process may often conflict with the goal of maintaining optimal operation, particularly when the system converges to locally optimal solutions without sufficient exploration. This fundamental challenge is well-documented in the field of adaptive robust MPC, where persistent excitation (PE) conditions are typically employed to ensure asymptotic identification of model parameters \cite{T10}. However, effectively integrating PE conditions with state and control constraints continues to pose significant practical challenges \cite{T11}. To address this issue, numerous algorithms have been developed that attempt to reconcile parameter identification and optimality tracking by introducing additional constraints into the receding horizon optimization framework of MPC. For instance, certain methods impose periodic input constraints to guarantee persistently exciting feedback laws. Nonetheless, such approaches frequently encounter difficulties in ensuring closed-loop stability or achieving convergence \cite{T12, T13}. Recent advancements in adaptive robust tube MPC formulations for linear systems \cite{T15,T14} have made progress by incorporating PE measures into the performance objective and including linearized PE conditions as constraints, thereby achieving recursive feasibility and input-to-state stability. However, these methods primarily focus on parameter identification under PE conditions, without fundamentally resolving the deeper conflict between parameter identification and optimality tracking within the auto-optimization framework.

To mitigate this conflict between estimation and tracking, a dual control approach has been proposed in \cite{Chen2}, which ensures that each control action simultaneously fulfills two purposes: driving the system toward the presumed optimal point (exploitation) while actively probing the environment to reduce uncertainty (exploration). This concept exhibits strong parallels with mechanisms in reinforcement learning, where emphasis is placed on the continuous interaction between the system and its environment to achieve an optimal balance between exploration and exploitation \cite{LiZ,Chen2,Chen,T28,tan2024multistep}.

\subsection{Contributions}
This paper introduces an auto-optimal MPC framework designed for uncertain environments, structured around two core methods: Exploitation-Oriented MPC (EO-MPC) and its extension, Active Learning MPC (AL-MPC). The framework addresses critical challenges in set-point tracking and parameter estimation, offering innovative solutions to achieve optimal performance in dynamic and uncertain environments.

The EO-MPC tackles the persistent excitation (PE) challenge by strategically incorporating virtual signals into the terminal set. This approach ensures parameter convergence without introducing nonconvex constraints on control inputs, a significant improvement over traditional methods that rely on restrictive PE assumptions or gradient descent techniques \cite{LiZ}. By maintaining recursive feasibility and guaranteeing parameter estimate convergence, EO-MPC provides a robust and computationally efficient foundation for the framework.

Building on EO-MPC, the AL-MPC  introduces a dynamic active learning mechanism that enables the system to progressively learn environmental characteristics and approach true optimal operating conditions. This approach is inspired by reinforcement learning paradigms, where continuous interaction between the system and its environment drives performance improvement \cite{Chen}. Unlike passive traditional methods\cite{LiZ,W4,kofinas2017reinforcement, W1}, AL-MPC actively quantifies environmental uncertainty using set-based parameter estimation, effectively balancing exploration and exploitation.

The auto-optimal MPC framework is characterized by four key innovations:
\begin{enumerate}
    \item The EO-MPC component addresses the PE challenge through terminal set modifications, ensuring parameter convergence without complex constraint additions. This foundational innovation enables the subsequent development of more sophisticated learning capabilities in AL-MPC.

     \item AL-MPC employs a dynamic MPC strategy that continuously adjusts to environmental changes, incorporating uncertainties into the objective function to optimize control actions in real-time.

\item AL-MPC's active exploration mechanism utilizes set-based parameter estimation to quantify uncertainty online, enabling the system to balance exploration and exploitation effectively.

\item The framework achieves rapid adaptation with minimal data, making it a practical solution for real-time systems requiring fast and reliable performance.
\end{enumerate}

\subsection{Notation}
 The subscript $\{i|k\}$ represents $i$ steps ahead at time step $k$ with $\{0|k\}=\{k\}$, $\mathbb{N}$ denotes the set of non-negative integers, the set $\mathbb{N}_{[0,N]}$ denotes the set of integers $\{0, 1, \cdots, N\}$, and  $I_n\in \mathbb{{R}}^{n\times n}$ denotes the identity matrix.  Consider $x\in \mathbb{R}^{n_x}$ and $u\in \mathbb{R}^{n_u}$, for a set $\mathbb{Z}\subseteq \mathbb{R}^{n_x+n_u}$, the projection operation is defined as $\text{Proj}_x(\mathbb{Z})=\{x\in\mathbf{R}^{n_x}, \exists u\in \mathbb{R}^{n_u},  (x,u)\in \mathbb{Z} \}$.

\section{Problem Description}
\subsection{System dynamics and constraints}
We consider a single-input and output autoregressive process with exogenous inputs (ARX)
\begin{equation}\label{eq101}
y_k = -\sum_{i=1}^{n} a_i\, y_{k-i} + \sum_{j=1}^{m} b_j\, u_{k-j},
\end{equation}
where $a_i$ and $b_j$ are the autoregressive and exogenous-input coefficients, and $n,m$ are their respective orders.

For prediction and control design, we introduce an equivalent state-space realization by defining the state as a vector of past measured signals. Specifically,
\begin{equation}\label{eq:x_def}
x_k =\! \begin{bmatrix}
y_{k-1} & \cdots & y_{k-n} & u_{k-1} & \cdots & u_{k-m}
\end{bmatrix}^T \in \mathbb{R}^{n_x},
\end{equation}
so that all components of $x_k$ are available at sampling time $k$ without an observer (they are past outputs and applied inputs).

With the definition \eqref{eq:x_def}, the ARX model \eqref{eq101} admits the standard lifted state-space form
\begin{align}
x_{k+1} &= A\,x_k + B\,u_k, \label{eq1a}\\
y_k     &= C\,x_k, \label{eq1b}
\end{align}
where $A,B,C$ are the companion/shift matrices induced by $\{a_i\}$ and $\{b_j\}$. In particular, $A$ shifts the output- and input-history stacks and inserts the linear combination $\{-a_i\}$ in the last output-history row, $B$ injects the current input $u_k$ into the last input-history position, and $C$ selects the first output-history entry so that $y_k=Cx_k$.
We assume the pair $(A,B)$ is controllable and $(A,C)$ observable; these hold for minimal ARX realizations under standard conditions.

In this paper we assume that $\mathbb{Z}$ contains no state-input coupling.
Specifically, $\mathbb{Z}$ factorizes as the Cartesian product
\begin{align}
\mathbb{Z} \;=\; \mathbb{X}\times\mathbb{U}
\;=\; \Big\{(x,u)\;\Big|\;
\underbrace{\begin{bmatrix} H_x & 0 \\[2pt] 0 & H_u \end{bmatrix}}_{=:H_{zu}}
\begin{bmatrix} x \\ u \end{bmatrix}
\;\le\;
\begin{bmatrix} h_x \\ h_u \end{bmatrix}
\Big\}, \label{eqZprod}
\end{align}
with
\begin{align}
\mathbb{X} &= \{\,x \in \mathbb{R}^{n_x} : H_x x \le h_x\,\}, \\
\mathbb{U} &= \{\,u \in \mathbb{R}^{n_u} : H_u u \le h_u\,\}.
\end{align}
Equivalently, there are no mixed inequalities involving both $x$ and $u$.
Under \eqref{eqZprod}, the projections satisfy $\mathrm{Proj}_x(\mathbb{Z})=\mathbb{X}$ and $\mathrm{Proj}_u(\mathbb{Z})=\mathbb{U}$ without loss of information.

\begin{remark}
Since the output is a linear function of the constructed state, bounds can be imposed equivalently on the state or on the output. State constraints induce output constraints, and output constraints can be enforced at the state level. The translation is lossless when the output map has full row rank; otherwise use exact projection or tight inner/outer approximations.
\end{remark}
\subsection{The unknown environment}

Consider a scenario in which the system (\ref{eq1a})-(\ref{eq1b}) operates in an unknown environment. The measurement model for the environment is given by
\begin{align}
    z_k = \phi^T(y_k)\theta^* + v_k, \label{eq7}
\end{align}
where $\theta^* \in \mathbb{R}^p$ denotes the unknown parameters that are specific to the operational environment, and $\phi^T(y_k) \in \mathbb{R}^p$ represents a bounded, measurement $z_k$ is used to identify the unknown parameter vector $\theta^*$. The term $v_k$ accounts for measurement noise.

\begin{remark}
Equation~\eqref{eq7} is introduced to quantify the environment via a noisy, linear-in-parameters regression evaluated on features of the plant output. In this way the environment signal is explicitly influenced by the system output. The unknown parameters summarize environment-specific preferences or costs and are identified online.
\end{remark}

For analytical simplicity, we assume that the function $\phi^T(y_k)\theta^*$ exhibits a unique extremum at the optimal condition $y_k = r^*$.  $\phi^T(y_k)\theta^*$ originates from either first-principle modeling, which often takes the form of quadratic or higher-order polynomial expressions, encapsulating key engineering objectives. Under this framework, the optimal condition $r^*$ can be explicitly expressed as a function of the unknown parameters, i.e., \( r^* = f(\theta^*) \).

\begin{assum}
The measurement noise $v_k$ is bounded as
\begin{align}
\mathbb{V} = \left\{v_k \in \mathbb{R} : H_v v_k \leq h_v\right\}. \label{eq8}
\end{align}
\end{assum}

\subsection{Control objective and challenges}

The primary objective of auto-optimal model predictive control (MPC) is to drive the system output \( y_k \) towards the unknown optimal operating point \( r^* = f(\theta^*) \). The controller aims to not only reach this optimal condition but also maintain system stability and performance despite external disturbances. Additionally, as the operational environment evolves, the auto-optimal MPC dynamically adapts, recalibrating the optimal operating point in real-time to ensure continued performance at peak efficiency. This adaptive capability allows the system to consistently track and adjust to changes, minimizing performance degradation and enhancing overall robustness.

If the environmental parameters are known, set-point tracking MPC aims to minimize the deviation between the actual output \( y_k \) and the optimal condition \( r^* \), with the objective of asymptotically driving this error to zero. For asymptotic stability to be achieved, the optimal condition \( r^* \) must correspond to an admissible equilibrium point \( (x^s, u^s) \in \mathbb{Z} \).When the condition is satisfied, the optimal condition \( r^* \) is considered reachable, and the tracking control objective can be realized. However, if this condition is not satisfied, stabilization of the system at the optimal condition \( r^* = f(\theta^*) \) becomes unachievable, thereby rendering the tracking control objective infeasible. Therefore, the optimal condition \( r^* \) must satisfy
\begin{equation}
\begin{bmatrix}
A - I_{n_x} & B \\
C & 0
\end{bmatrix}
\begin{bmatrix}
x^s \\
u^s
\end{bmatrix}
=
\begin{bmatrix}
0 \\
r^*
\end{bmatrix},
\label{eq10}
\end{equation}
with
\begin{equation}
\text{rank}\left(
\begin{bmatrix}
A - I_{n_x} & B \\
C & 0
\end{bmatrix}
\right) = n_x + n_y.
\label{eq11}
\end{equation}
The rank condition guarantees the existence of a solution to the equilibrium equations \cite{dos2024set,ferramosca2010mpc}.

Assuming that condition (\ref{eq11}) holds, we define the sets of steady states \( \mathbb{Z}_s \) and reachable outputs \( \mathbb{Y}_r \) as follows
\begin{align}
&\mathbb{Z}_s = \left\{ (x^s, u^s) \ \big| \ x^s = A x^s + B u^s, \ (x^s, u^s) \in \mathbb{Z} \right\}, \label{eq12} \\
&\mathbb{Y}_r = \left\{ r \ \big| \ r = C x^s, \ (x^s, u^s) \in \mathbb{Z}_s \right\}. \label{eq13}
\end{align}
Starting from an initial state \( x_0 \), the MPC optimization problem is formulated as
\begin{subequations}\label{eq14}
\begin{align}
\min_{\mathbf{u}_k}  & \sum_{i=0}^{N-1} \left( \|x_{i|k} - x^s\|_Q^2 + \|u_{i|k} - u^s\|_R^2 \right) + \|x_{N|k} - x^s\|_S^2, \nonumber\\
\text{s.t} \quad
& x_{i+1|k} = A x_{i|k} + B u_{i|k}, \quad \forall i \in \mathbb{N}_{[0, N-1]}, \\
& (8), (x_{i|k}, u_{i|k}) \in \mathbb{Z}, \quad \forall i \in \mathbb{N}_{[0, N-1]}, \\
& x_{N|k} \in \mathbb{X}_f(x^s),
\end{align}
\end{subequations}
 where \( x_{i|k} \) and \( u_{i|k} \) represent the predicted \( i \)-th  step state and control input at time step $k$. \( \mathbf{u}_k = [u_{0|k}, u_{1|k}, \ldots, u_{N-1|k}] \) is the control policy. The terminal constraint \( \mathbb{X}_f(x^s) = \mathbb{X}_f \oplus x^s = \{ x \mid H_f (x - x^s) \leq h_f \} \) is contingent upon the equilibrium state point  \( x^s \). \( N \) defines the finite prediction horizon,  the matrices \( Q \), \( R \), and \( S \) are positive definite, ensuring the quadratic nature of the cost function.

In above MPC optimization problem, the following issues arise:

\begin{enumerate}
    \item \textit{Unknown Optimal Condition:} The optimal tracking condition \( r^* \) is unknown and varies with changes in the unknown parameter \( \theta^* \). Since the environmental parameters are unknown, the optimal condition cannot be directly obtained and must be estimated and tracked through the system control input.

    \item \textit{Persistent Excitation Condition for Control Input:} Estimating the unknown parameter \( \theta^* \) requires that the control input satisfies the persistent excitation (PE) condition. Only when the control input signals sufficiently excite the system can accurate parameter estimation be ensured; otherwise, estimation errors may arise, affecting system performance and stability \cite{T15,T16,28}.

    \item \textit{Challenges in Recursive Feasibility and Convergence Analysis:}
Auto-optimal MPC may face feasibility issues if parameter estimates deviate from actual values or if control input lacks sufficient excitation over time. This can disrupt recursive feasibility, preventing feasible solutions at each step. Inaccurate estimates or insufficient excitation may also hinder system convergence to the desired point, compromising control objectives \cite{W1, limon2015mpc}.
\end{enumerate}

\section{Exploitation-Oriented MPC}

The exploitation-oriented MPC (EO-MPC) forms the foundation for active learning MPC (AL-MPC), which is further developed in subsequent sections. This section introduces the EO-MPC approach, which begins with robust set-based parameter estimation, emphasizing the importance of the PE condition in achieving accurate parameter identification. A cost function and terminal constraint are then proposed to guide the operation of the EO-MPC. The section concludes with proofs of recursive feasibility and convergence, thereby establishing the theoretical soundness of the proposed approach.

\subsection{Robust set-based parameter estimation}
Robust set-based parameter estimation techniques are extensively used in adaptive control theory. As shown in \cite{T15,T16}, when the regressor satisfies the PE condition, robust set-based parameter estimation converges to the true parameter values with probability 1.

Consider an information state ${\bf I}_k$,  defined as
\begin{equation}\label{eq15}
{\bf I}_k=\left[z_0,z_{1},\cdots,z_{k},u_{0},u_{1},\cdots,u_{k-1}\right],
\end{equation}
starting with ${\bf I}_0 = \left[z_0\right]$. The information state ${\bf I}_k$ facilitates the recursive refinement of the parameter set $\Theta_{k}$, crucial for achieving the desired control strategy. Formally, the estimated set of unknown parameters  $\theta^*$ is defined as

\begin{equation}
    \Theta_k = \{\hat{\theta}_k : H_{\theta_k}\hat{\theta}_k \leq h_{\theta_k} | {\bf I}_k\},
\end{equation}
where $\hat{\theta}_k$ represents the estimated parameter at time step $k$, and constraint matrix $H_{\theta_k}$ and vector $h_{\theta_k}$ can iteratively change according to Eq.(\ref{eq18}). According to (\ref{eq8}),  the information state ${\bf I}_k$ updates $\Theta_k$ by constructing a non-falsified parameter set
\begin{equation}\label{eq17}
    \Delta_k = \{\hat{\theta}_k : H_{\Delta_k}\hat{\theta}_k \leq h_{\Delta_k} | {\bf I}_k\},
\end{equation}
where $H_{\Delta_k} = -H_v\phi^T(y_k)$ and $h_{\Delta_k} = h_v - H_v z_k$. The parameter set $\Theta_k$ is refined by intersecting with the non-falsified set $\Delta_k$:
\begin{equation}\label{eq18}
    \Theta_k = \Theta_{k-1} \bigcap \Delta_k.
\end{equation}
This intersection ensures that the updated set $\Theta_k$ does not exceed the bounds established by the preceding set $\Theta_{k-1}$, maintaining the set's monotonicity.

\subsection{Persistent excitation}
The regressor $\phi(y_k)$ in Eq. (\ref{eq7}) is said to be persistently exciting if it satisfies the following condition over a finite horizon $N_u$
\begin{equation}\label{eq19}
    \sum_{k}^{k+N_u-1} \mathbb{E}[\phi(y_k) \phi^T(y_k)] \succeq \beta I_p,
\end{equation}
where $\beta$ is a positive scalar, and $I_p$ is the identity matrix of appropriate dimension. The interval $[k, k+1, \dots, k+N_u-1]$ is referred to as the PE window.

\begin{lemma}
Under Assumption 1, if the regressor $\phi(y_k)$ satisfies the PE condition for Eq. (\ref{eq19}), the estimated parameter set $\Theta_k$ converges to the true parameter set $\{\theta^*\}$ as $k \to \infty$ with probability 1.
\end{lemma}
The Lemma 1 is derived based on the rigorous analysis presented in Theorem 3 and Corollary 3 of \cite{T15}.

A widely adopted method to ensure the PE condition in dynamical systems involves incorporating a virtual perturbation signal into the input channel \cite{T15,T16}. The composite control input can be expressed as
  $u_k = u_{d,k} + s_k$, where \( u_{d,k} \) represents the nominal control signal, and $s_k$ denotes a virtual perturbation signal with known stochastic properties.

\begin{assum}
The virtual signal $s_k \in \mathbb{S}$ satisfies
\begin{equation}
\mathbb{E}(s_k) = 0 \quad \text{and} \quad \mathbb{E}(s_k s_k^T) = \epsilon_s I_{n_u},
\end{equation}
where $\epsilon_s > 0$ quantifies the perturbation intensity, and $\mathbb{S}$ is a known polytopic bounded set, and the sequence $\{s_k\}$ forms a zero-mean independent and identically distributed (i.i.d.) process.
\end{assum}

Incorporating the  composite control input and the system model (\ref{eq101}), the augmented output dynamics become
\begin{equation}\label{eq1011}
\hat{y}_k = \underbrace{-\sum_{i=1}^{n} a_i y_{k-i} + \sum_{j=1}^{m} b_j u_{d,k-j}}_{y_k} + \underbrace{\sum_{j=1}^{m} b_j s_{k-j}}_{\tilde{s}_k}.
\end{equation}

\begin{assum}
The effective perturbation term  $\tilde{s}_k$ inherits statistical properties
\begin{equation}
\mathbb{E}(\tilde{s}_k) = 0 \quad \text{and} \quad \mathbb{E}(\tilde{s}_k \tilde{s}_k^T) = \epsilon_{\tilde{s}} I_{n_u},
\end{equation}
where $\epsilon_{\tilde{s}}>0$. Furthermore, $\tilde{s}_k$ is statistically independent of $y_k$, and the nonlinear feature mapping
$\phi(\hat{y}_k) = \left [ \hat{y}_k, \hat{y}_k^2,\cdots,\hat{y}_k^n\right ]^T$
 guarantees
$\mathbb{E}\big[\phi(\hat{y}_k)\phi^T(\hat{y}_k)\big] \succ 0$.

\end{assum}

For the quadratic regressor \(\phi(\hat{y}_k) = \begin{bmatrix}\hat{y}_k & \hat{y}_k^2\end{bmatrix}^T\), substituting  the central moments of \(\tilde{s}_k\) into the expectation matrix yields
\begin{align}\label{eq10113}
\mathbb{E} \left[ \phi(\hat{y}_k) \phi^T(\hat{y}_k) \right]
= \begin{bmatrix}
y_k^2 + \epsilon_{\tilde{s}} & y_k^3 +3y_k\epsilon_{\tilde{s}} \\
y_k^3 +3y_k\epsilon_{\tilde{s}} & y_k^4 + 6y_k^2\epsilon_{\tilde{s}} + 3\epsilon_{\tilde{s}}^2
\end{bmatrix}.
\end{align}
Since both diagonal entries are positive and the determinant,
\begin{align}
\det \left(\mathbb{E} \left[ \phi(\hat{y}_k) \phi^T(\hat{y}_k) \right] \right)
&= y_k^4\epsilon_{\tilde{s}} + 3\epsilon_{\tilde{s}}^3 > 0.
\end{align}
According to Sylvester's criterion, it follows that
$\mathbb{E} \left[ \phi(\hat{y}_k) \phi^T(\hat{y}_k) \right] \succ 0.$
For the third regressor \(\phi(\hat{y}_k) = \begin{bmatrix}\hat{y}_k & \hat{y}_k^2 & \hat{y}_k^3\end{bmatrix}^T\), a similar analysis demonstrates the positive definiteness of \(\mathbb{E} \left[ \phi(\hat{y}_k) \phi^T(\hat{y}_k) \right]\) via the following nonzero determinant:
\begin{align}\label{eq10114}
\det \left(\mathbb{E} \left[ \phi(\hat{y}_k) \phi^T(\hat{y}_k) \right] \right)
&= 10y_k^{10}\epsilon_{\tilde{s}} + 25y_k^8\epsilon_{\tilde{s}}^2 + 167y_k^6\epsilon_{\tilde{s}}^3 \nonumber\\
&+ 144y_k^4\epsilon_{\tilde{s}}^4 + 72y_k^2\epsilon_{\tilde{s}}^5 + 18\epsilon_{\tilde{s}}^6 > 0.\nonumber
\end{align}
\textbf{Remark 1.} \textit{Although the proposed theoretical framework can be extended to a general \(n\)-th order regressor, when \(n \geq 4\), verifying the positive definiteness condition
$\mathbb{E} \left[ \phi(\hat{y}_k) \phi^T(\hat{y}_k) \right] \succ 0$
analytically becomes computationally challenging. This is due to the fact that the order of the joint moments required for the verification grows at a factorial rate \(\mathcal{O}((2n)!)\), leading to a dramatic increase in computational complexity. Nevertheless, the analytical results established for the second- and third-order cases  are sufficient to ensure that the fundamental assumptions for applying the theoretical framework hold true.}

By Assumption~2, the full-rank property propagates temporally, guaranteeing existence of  \( \beta>0 \) such that the accumulated correlation matrix
\begin{equation}
R_k =\sum_{k}^{k+N_u-1} \mathbb{E} \left[ \phi(\hat{y}_k) \phi^T(\hat{y}_k) \right] I_p \succeq \beta I_p.
\end{equation}
satisfies the PE condition in (\ref{eq19}).

In auto-optimal MPC, the relation  $r^* = f(\theta^*)$  shows how the optimal operational condition depends on the environmental parameter  $\theta^*$. Given that the estimated parameters $\hat{\theta}_k$ reside within the set $\Theta_k$, the estimated operational condition  $\hat{r}_k$ can be determined as follows
\begin{equation}\label{eq20}
   \hat{r}_k = f(\hat{\theta}_k), \ \forall \hat{\theta}_k \in \Theta_{k}.
\end{equation}
\subsection{Cost function}
The primary objective is to drive the system output towards the estimated parameter $\hat{r}_k=f(\hat{\theta}_k)$. Given the stochastic nature of $\hat{r}_k$, designing an appropriate cost function is pivotal for guiding the system towards its goal. The cost function is formulated as follows
\begin{align}
&J(x_k, {\bf u}_k)=\mathbb{E}\left\{\sum_{i=0}^{N-1}\left(||x_{i|k}-\hat{x}_k^s||_{Q}^2 +||u_{i|k}-\hat{u}_k^s||_{R}^2\right)\right.\nonumber\\
&\left. \ \ \ \ \ \ \ \ \ +||x_{N|k}-\hat{x}_k^s||_{S}^2\Big|{\bf I}_k \right\}
\end{align}
 where the estimated equilibrium point $(\hat{x}_k^s, \hat{u}_k^s)\in\mathbb{Z}$  must satisfy
\begin{equation}
\left[\begin{array}{cc}
A-I_{n_x}& B\\
C &0
\end{array}\right]
\left[\begin{array}{c}
\hat{x}_k^s\\
\hat{u}_k^s
\end{array}\right] =
\left[\begin{array}{c}
0\\
\hat{r}_k
\end{array}\right]. \label{T23}
\end{equation}

The terminal weight matrix $S$ and the control gain $K$ are selected based on the solution of the following Lyapunov equation
$(A+BK)^TS(A+BK)-S=-Q-K^TRK$.

Furthermore, the determination of the estimated optimal operational condition  involves computing the mean and covariance of the estimated optimal reference condition, denoted as $\bar{r}_k=\mathbb{E}[\hat{r}_k^s\big|{\bf I}_k ]$, and $P_k^r=\mathbb{E}[(\hat{r}_k-\bar{r}_k)(\hat{r}_k-\bar{r}_k)^T\big|{\bf I}_k ]$ respectively. Similarly, the mean vectors of the steady state and input estimates are represented as $\bar{x}_k^s=\mathbb{E}[\hat{x}_k^s\big|{\bf I}_k]$, $\bar{u}_k^s=\mathbb{E}[\hat{u}_k^s\big|{\bf I}_k]$, and their covariances are $P_k^x=\mathbb{E}[(\hat{x}_k^s-\bar{x}_k^s)(\hat{x}_k^s-\bar{x}_k^s)^T\big|{\bf I}_k]$, $P_k^u=\mathbb{E}[(\hat{u}_k^s-\bar{u}_k^s)(\hat{u}_k^s-\bar{u}_k^s)^T\big|{\bf I}_k]$.

The cost function can be further decomposed as follows
\begin{eqnarray}
 &&J(x_k, {\bf u}_k)\nonumber\\
  &&=\sum_{i=0}^{N-1}\left(||x_{i|k}-\bar{x}_k^s||_{Q}^2 +||u_{i|k}-\bar{u}_k^s||_{R}^2\right)+||x_{N|k}-\bar{x}_k^s||_{S}^2\nonumber\\
 &&+(N-1)[\text{Trace}(QP_k^x+RP_k^u)]+\text{Trace}(SP_k^x).
\end{eqnarray}
Since both $P_k^x$ and $P_k^u$ are constant and known at time step $k$, the term  $(N-1)[\text{Trace}(QP_k^x+RP_k^u)]+\text{Trace}(SP_k^x)$ can be omitted from the cost function.
\subsection{Terminal constraint set design}
 To ensure that the terminal constraint set must remain within the state constraint set,  a key aspect of an method in \cite{28,Rao} involves introducing a scalar variable $\alpha_k$, which dynamically scales the terminal constraint set $\mathbb{X}_f$ around the estimated state $\bar{x}_k^s$:
\begin{align}
 x_{N|k}\in \alpha_k \mathbb{X}_f(\bar{x}_k^s)\subseteq \mathbb{X}.
\end{align}
where $\mathbb{X}_f(\bar{x}_k^s)=\mathbb{X}_f\oplus\bar{x}_k^s=\{x: H_f(x-\bar{x}_k^s)\leq h_f\}$  depends on $\bar{x}_k^s$. The scalar variable $\alpha_k$ is introduced to adjust the size of the terminal constraint set, ensuring it remains within the state constraint set $\mathbb{X}$ while accommodating changes in the estimated steady state.

\begin{lemma}
Let \( \mathbb{X}_f(x^s) \) be an invariant set for the system described by $
x_{k+1} - x^s = (A + BK)(x_k - x^s),$
meaning that if \( x_k \in \mathbb{X}_f(x^s) \), then \( x_{k+1} \in \mathbb{X}_f(x^s) \). For a scalar  \( 0 < \alpha_k \leq 1 \), the scaled set \( \alpha_k \mathbb{X}_f(x^s) \) remains an invariant set for the system.
\end{lemma}

The lemma follows directly from the scaling invariance of linear systems \cite{Blanchini}.

To ensure the PE condition, a common method involves adding virtual signal to the control input.
However, this approach can lead to poor tracking performance and may violate constraints on the states and control inputs.
One effective way to mitigate these undesirable effects is to incorporate a terminal horizon control law with virtual signal $s_k$, as demonstrated in \cite{T15}. The terminal horizon control law is expressed as
$u_{i|k} = K(x_{i|k} - \bar{x}_k^s) + \bar{u}_k^s + s_{i|k}, \quad i \in \mathbb{N}_{[N-1,N+N_u+1]}$.

Due to the terminal horizon control law with the virtual signal $s_k$, the  terminal constraint set \(\mathbb{X}_f\) need satisfy the following three conditions for a given gain \(K\)

\begin{itemize}
    \item $\alpha_k \mathbb{X}_f(\bar{x}_k^s)\in \mathbb{X}, \forall (\bar{x}_k^s, \bar{u}_k^s)\in \mathbb{Z}$;
    \item \(\forall x_{N|k} \in \alpha_k \mathbb{X}_f(\bar{x}_k^s)\), it holds that $(x_{N|k}, u_{N|k}) \in \mathbb{Z}, \forall  s_{N|k}\in \mathbb{S}$;
    \item The dynamics are constrained such that $(A+BK)\mathbb{X}_f(\bar{x}_k^s)\oplus B\mathbb{S} \in \alpha_k \mathbb{X}_f(\bar{x}_k^s).$
\end{itemize}
The equilibrium point \((\bar{x}_k^s, \bar{u}_k^s)\), which is computed based on the optimal operating condition \(\bar{r}_k\), may not always satisfy  \((\bar{x}_k^s, \bar{u}_k^s)\in \mathbb{Z}\). To resolve this issue, a practical approach is to select the equilibrium point, described by the following convex optimization problem
\begin{subequations}
\begin{align}
&\bar{r}_k^s, \bar{x}_k^s, \bar{u}_k^s = \arg \min_{\bar{r}_k^s, \bar{x}_k^s, \bar{u}_k^s} ||\bar{r}_k^s - \bar{r}_k||_D^2,\label{TT1} \\
&\left[\begin{array}{cc}
A - I_{n_x} & B \\
C & 0
\end{array}\right]
\left[\begin{array}{c}
\bar{x}_k^s \\
\bar{u}_k^s
\end{array}\right] =
\left[\begin{array}{c}
0 \\
\bar{r}_k^s
\end{array}\right], (\bar{x}_k^s, \bar{u}_k^s) \in \mathbb{Z},\label{TT3}
\end{align}
\end{subequations}
where $D$ is positive matrix, \(\bar{r}_k^s\) is introduced as an artificial output and serves as an additional decision variable in the optimization problem. This ensures that the equilibrium point \((\bar{x}_k^s, \bar{u}_k^s)\in \mathbb{Z}\).
This approach helps prevent the loss of feasibility in cases where the equilibrium point does not lie within the constraint set.

\subsection{Optimization problem}
Following the presented penalty method (\ref{TT1})-(\ref{TT3}), the proposed controller is based on the solution at each sampling time of an optimal control problem.  The EO-MPC tracking optimal problem can be stated as follows
\begin{subequations}\label{eq26}
\begin{align}
&\!\!\!\!\!\!\!\!\!\!\!  \min_{\alpha_k, \mathbf{u}_k,\bar{r}_k^s,\bar{x}_k^s,\bar{u}_k^s}\ J(x_k, {\bf u}_k)+||\bar{r}_k^s-\bar{r}_k||_D^2 \\
\!\!\!\!\!\!\!\text{s.t.}~& x_{i+1|k}=Ax_{i|k}+Bu_{i|k},(31b),\\
\!\!\!\!\!\!\!&(\bar{x}_k^s,\bar{u}_k^s), (u_{i|k}, x_{i|k})\in \mathbb{Z},x_{N|k}\in\alpha_k\mathbb{X}_f(\bar{x}_k^s),
\end{align}
\end{subequations}
where $J(x_k, {\bf u}_k)$ is given by Eq. (28).  The second term of cost function offset cost functional penalizes the deviation of the artificial steady-output \(\bar{r}_k^s\) from the target output \(\bar{r}_k\), ensuring that the system converges towards the desired operating condition.

\subsection{Persistence of excitation check}

The PE condition in Eq. (\ref{eq19}) is non-convex and involves quadratic matrix inequalities. It creates a challenge in solving the corresponding MPC problem when directly incorporating  it as a constraint in the optimization. However, it can be addressed using a sample-based verification method in \cite{T16} applied to the solution of the
EO-MPC optimization. This method compares the obtained solution with a reference solution known to satisfy the PE condition on average. If the sample-based check fails, the reference solution is used as a fallback.

Given the solution \({\bf u}_k\) from the EO-MPC optimization at time step \(k\), we evaluate the persistence of excitation for \({\bf u}_k^*\) by comparing it with a reference solution \(\hat{{\bf u}}_k\). The reference solution \(\hat{{\bf u}}_k = [\hat{u}_{0|k}, \hat{u}_{1|k}, \dots, \hat{u}_{N-1|k}]\) is derived from the previous time step's solution \(\hat{{\bf u}}_{k-1}\) and includes the virtual signal $s_{i|k}$ into the system:

\begin{align}
 \hat{u}_{i|k}=\left\{
        \begin{aligned}
        &u_{i+1|k-1}^*, &i \in \mathbb{N}_{[0,N-2]},\\
        &K(x_{i|k}-\bar{x}_k^s)+\bar{u}_k^s+s_{i|k}, &i \in \mathbb{N}_{[N-1,N_u-1]}.
        \end{aligned}\nonumber
        \right.
\end{align}
To ensure that the PE condition extends beyond the first \(N\) time steps of the prediction horizon, we redefine the control input for \(i \geq N-1\) as \(u_{i|k}=K(x_{i|k}-\bar{x}_k^s)+\bar{u}_k^s+s_{i|k}\), and modify the PE matrix as:

\begin{align}
    &R_{i|k}({\bf u}_k)=\sum_{j=i}^{i+N-1} \phi(y_{j|k}) \phi^T(y_{j|k}) \nonumber\\
    &\ \ \ \ \ \ \ \ \ \ \ \ \ \ \ \ \ \ + \sum_{j=i+N}^{i+N_u-1} \mathbb{E}[\phi(y_{j|k}) \phi^T(y_{j|k})].
\end{align}
As stated in Section 3.2, when the virtual signal is used, there exists a constant \(\hat{\beta}\) such that: $R_{i|k}(\hat{{\bf u}}_k) \succeq \hat{\beta} I_p$.
To verify whether the optimized control input ${\bf u}_k^*$ obtained from (\ref{eq26}) is positive definite, we compute \(\beta_{i|k}\) as:
\begin{equation}
   \beta_{i|k}=\max_{\beta \in \mathbb{R}} \beta \ \text{s.t.} \ R_{i|k}({\bf u}_k^*) \succeq \beta I_p.
\end{equation}
The check is considered to have failed if \(\beta_{i|k} \leq 0\), and in this case, \({\bf u}_k^*\) is redefined by setting it to \(\hat{{\bf u}}_k\).

\subsection{Recursive feasibility and Convergence}
The theorems below articulate the findings pertaining to recursive feasibility and convergence.
\begin{theorem} Suppose that Assumptions 1-4 hold, if the optimization problem (\ref{eq26}) is feasible at the time $k = 0$, then, the optimization problem (\ref{eq26}) will remain feasible for all $k>0$.
\end{theorem}
{\sc Proof:} For a feasible state $x_k\in \mathbb{X}$ at time step $k$, optimal control sequence of the optimization problem (\ref{eq26}) is given by
${\bf u}_k^*=[u_{0|k}^*,u_{1|k}^*,\cdots,u_{N-1|k}^*],$
with the optimal predicted state sequence ${\bf x}_k^*=[x_{0|k}^*,x_{1|k}^*,\cdots,x_{N-1|k}^*,x_{N|k}^*],$
and $\alpha_k^*$, $\bar{r}_k^{s*}$, $\bar{x}_k^{s*}$, $\bar{u}_k^{s*}$.  Defining an auxiliary feasible input sequence
${\bf u}_{k+1}=[u_{1|k}^*,u_{2|k}^*,\cdots,u_{N-1|k}^*,K(x_{N|k}^*-\bar{x}_k^{s*})+\bar{u}_k^{s*}+s_{N|k}],$
${\bf x}_{k+1}=[x_{1|k}^*,x_{2|k}^*,\cdots,x_{N|k}^*,(A+BK)(x_{N|k}^*-\bar{x}_k^{s*})+\bar{x}_k^{s*}+Bs_{N|k}],$ and maintaining $\alpha_{k+1}=\alpha_k^*$, $\bar{r}_{k+1}^s=\bar{r}_k^{s*}$, $\bar{x}_k^{s*}$, $\bar{u}_k^{s*}$. According to the designed terminal set $\alpha_k^*\mathbb{X}_f(\bar{x}_k^{s*})$ which satisfies condition $(A+BK)\mathbb{X}_f(\bar{x}_k^s)\oplus B\mathbb{S} \in \alpha_k \mathbb{X}_f(\bar{x}_k^s)$,  it follows that $x_{N|k+1}=(A+BK)(x_{N|k}^*-\bar{x}_k^{s*})+\bar{x}_k^{s*}+Bs_{N|k}\in\alpha_k^*\mathbb{X}_f(\bar{x}_k^{s*})$ proving that the closed-loop system is recursively feasible.

The following lemma plays an important role on proof of the convergence and follows directly from \cite{monasterios2018model}.
\begin{lemma}
If $f(\cdot)$ is a continuous function, then there exists a $\mathcal{K}_\infty$ function $\alpha(\cdot)$ such that $|f(x)-f(y)|\leq \alpha(|x-y|)$, for all $x\in C$ and $y\in D$ with $C \subseteq D\subseteq \mathbb{R}^n$.
\end{lemma}

\begin{theorem}If the optimization problem (\ref{eq26}) is  feasible for all $k>0$, $y_k$ asymptotically converges to $r^*=f(\theta^*)$.
\end{theorem}

{\sc Proof:} Consider the following formulation:
\begin{align}
&\!\!\!\!\!\!\!\!\!  J_{k}=\sum_{i=0}^{N-1}\left(||x_{i|k}-\bar{x}_k^s||_Q^2 +||u_{i|k}-\bar{u}_k^s||_R^2\right) \nonumber\\
&+||x_{N|k}-\bar{x}_k^s||_{S}^2 +||\bar{r}_k^s-\bar{r}_k||_D^2
\end{align}
Employing the feasible, albeit possibly suboptimal, solution sequence:${\bf u}_{k+1}=[u_{1|k}^*,u_{2|k}^*,\cdots,u_{N-1|k}^*,K(x_{N|k}^*-\bar{x}_k^{s*})+\bar{u}_k^{s*}],$
${\bf x}_{k+1}=[x_{1|k}^*,x_{2|k}^*,\cdots,x_{N|k}^*,(A+BK)(x_{N|k}^*-\bar{x}_k^{s*})+\bar{x}_k^{s*}]$, and $\alpha_{k+1}=\alpha_k^*$, $\bar{r}_{k+1}^s=\bar{r}_k^{s*}$, $\bar{x}_k^{s*}$, $\bar{u}_k^{s*}$ at the next time $k + 1$,  $J_{k+1}$ satisfies
$J_{k+1}\leq J_k-\left(||x_{0|k}-\bar{x}_k^s||_{Q}^2 +||u_{0|k}-\bar{u}_k^s||_{R}^2\right)$.
This implies that $J_k$ is strictly decreasing provided  $x_{k}\neq\bar{x}_k^s$ and $u_{k}\neq\bar{u}_k^s$. Therefore,  $\lim_{k\rightarrow\infty} y_k\rightarrow \bar{r}_k$.
 Given that $\phi(y_k)$ is persistently exciting, the parameter estimate $\hat{\theta}$ converges to the singleton containing the true parameter $\theta^*$. Let
\begin{align}
& \tilde{J}_k = \sum_{i=0}^{N-1} \left( ||x_{i|k} - x^s||_Q^2 + ||u_{i|k} - u^s||_R^2 \right) \nonumber\\
&+ ||x_{N|k} - x^s||_{S}^2 + ||\bar{r}_k^s - r^*||_D^2
\end{align}
and $J_k$ have the same structure. Since both are continuous functions, it follows from Lemma 3 that there exists a $\mathcal{K}_\infty$ function $\alpha(\cdot)$ such that
\begin{equation}
J_k - \tilde{J}_k \leq \alpha\left( |\bar{x}_k^s - x^s| + |\bar{r}_k^s - r^*| \right),
\end{equation}
for all $x_k \in \mathbb{X}, \bar{x}_k^s \in \mathbb{X}$, and $x^s \in \mathbb{X}$.

Therefore, as $k \rightarrow \infty$, $\alpha(0) = 0$. Consequently, $\lim_{k \rightarrow \infty} \bar{r}_k \rightarrow f(\theta^*)$, establishing the asymptotic convergence of $y_k$ to $f(\theta^*)$.

\section{Active Learning MPC}

The EO-MPC algorithm is engineered to optimize control actions using the most recent and pertinent data, with a primary focus on enhancing immediate system performance. While this approach excels in leveraging known information to maximize short-term performance, it often exhibits limitations when faced with substantial future uncertainties. Specifically, the EO-MPC may underperform in adaptive learning contexts where the exploration of unknown parameters is critical for enhancing long-term system performance and adaptability. To mitigate these limitations and provide a holistic framework that balances the need for immediate exploitation with the imperative for exploration, we introduce the AL-MPC approach. This methodology synergizes exploitation of current knowledge with proactive exploration of unknown parameters, thereby presenting a robust strategy for managing and reducing uncertainty while maintaining optimal system performance.

The terminal constraint set design and the maintenance of persistent excitation in the AL-MPC method follow the strategies outlined in Section III, leveraging similar robust approaches to ensure system convergence.

\subsection{Predicted set-based parameter estimation}

The AL-MPC  uses  predicted future measurements $\bar{z}_{i|k}$, given by
\begin{align}
\bar{z}_{i|k}=\phi^T(y_{i|k})\bar{\theta}_{k}.
\end{align}
where $y_{i|k}$denotes the predicted output and $\bar{\theta}_{k}$ represents the current parameter estimate.
An predicted information state ${\bf I}_{i|k}$ is defined as
${\bf I}_{i|k}=\left[{\bf I}_k,u_{0|k},\bar{z}_{1|k},\cdots, u_{i|k},\bar{z}_{i|k}\right],$
where ${\bf I}_k$ denotes the information state at the current time step, and
$u_{i|k}$ represents the predicted control input.
Using these predicted information states, the predicted non-falsified set is defined as

\begin{align}\label{eq33}
\hat{\Delta}_{i|k}&=\{\hat{\theta}_{i|k}: H_v[\bar{z}_{i|k}-\phi^T(y_{i|k})\hat{\theta}_{i|k}]\leq h_v |{\bf I}_{i|k}\},\nonumber\\
&=\{\hat{\theta}_{i|k}: H_\Delta\hat{\theta}_{i|k}\leq h_{\Delta_{i|k}}|{\bf I}_{i|k}\}
\end{align}
The hyperplanes defining the sets $\hat{\Delta}_{i|k}$ depend affinely on the control input $u_{i|k}$ as seen in (\ref{eq33}).A sequence of predicted parameter sets can now be defined as
\begin{eqnarray}\label{eq43}
\hat{\Theta}_{i|k}=\{\hat{\theta}_{i|k}: H_\theta\hat{\theta}_{i|k}\leq h_{\theta_{i|k}} |I_{i|k}\}=\hat{\Delta}_{i|k}\cap \Theta_{k}.
\end{eqnarray}
An illustration of the parameter estimate, the predicted constraints, and parameter set is shown in Fig. 1.
\begin{figure}
\centering
\includegraphics*[width=7cm]{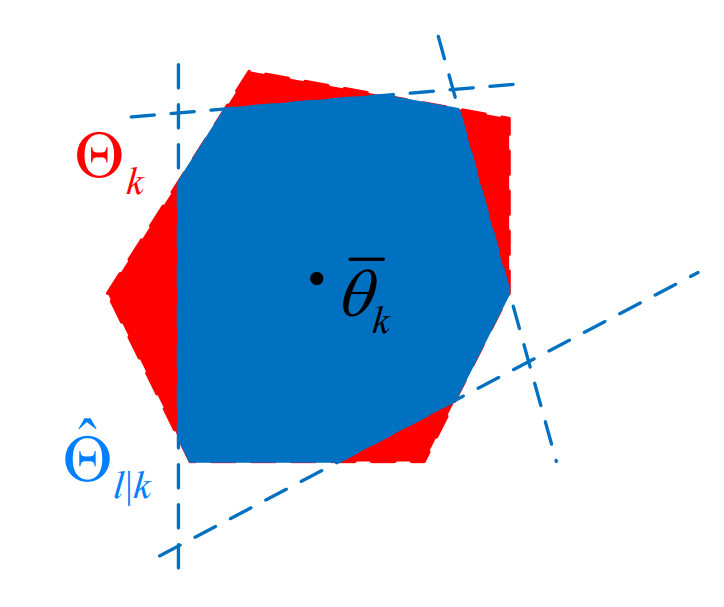}
\caption{ The parameter set $\Theta_{k}$ is bounded by the constraints
shown in red, and the parameter estimate $\bar{\theta}_k$ lies inside this set.
The predicted bounds are shown as dashed blue lines, which
depend on the control input variables $u_{i|k}$. The blue shaded
region shows the resulting predicted parameter set $\hat{\Theta}_{i|k}$.}
\end{figure}

\subsection{Cost function}
The cost function as the sum of stage costs and a terminal cost is defined as
\begin{align}\label{EqTY1}
 &J_{DC}(x_0, {\bf u}_k)=\mathbb{E}\left\{\sum_{i=0}^{N-1}\left(||x_{i|k}-\hat{x}_{i|k}^s||_{Q}^2 +||u_{i|k}-\hat{u}_{i|k}^s||_{R}^2\right)\right.\nonumber\\
 &\ \ \ \ \ \ \ \ \  \ \ \ \ \ \ \ \ \  +||x_{N|k}-\hat{x}_{N|k}^s||_{S}^2 |{\bf I}_{i|k}\Big\}
\end{align}
where $\hat{x}_{i|k}^s$ and $\hat{u}_{i|k}^s$ represent the estimated steady state, determined based on $\hat{r}_{i|k}$.

The predicted mean and covariance of the operational condition are denoted as  $\bar{r}_{i|k}=\mathbb{E}[\hat{r}_{i|k}^s|{\bf I}_{i|k}]$, and $P_{i|k}^r=\mathbb{E}[(\hat{r}_{i|k}-\bar{r}_{i|k})(\hat{r}_{i|k}-\bar{r}_{i|k})^T|{\bf I}_{i|k}]$ respectively.  The covariance of the steady-state and input estimates are represented as $P_{i|k}^x=\mathbb{E}[(\hat{x}_{i|k}^s-\bar{x}_{i|k}^s)(\hat{x}_{i|k}^s-\bar{x}_{i|k}^s)^T|{\bf I}_{i|k}]$, $P_{i|k}^u=\mathbb{E}[(\hat{u}_{i|k}^s-\bar{u}_{i|k}^s)(\hat{u}_{i|k}^s-\bar{u}_{i|k}^s)^T|{\bf I}_{i|k}]$. According to the mean and covariance calculation formula,  Eq.(\ref{EqTY1}) is  reformulated as  the sum of the following two components
\begin{equation}\label{TY23}
J_{DC}(x_k,\mathbf{u}_{k})= J_{ET}(x_k,\mathbf{u}_k)+J_{ER}(P_k^r,\mathbf{u}_{k})
\end{equation}
where
\begin{eqnarray}
&&J_{ET}(x_k,\mathbf{u}_k)=\sum_{i=0}^{N-1}\left(||x_{i|k}-\bar{x}_{i|k}^s||_{Q}^2 +||u_{i|k}-\bar{u}_{i|k}^s||_{R}^2\right)\nonumber\\
&&\ \ \ \ \ \  \ \ \ \ \ \ \ \ \ +||x_{N|k}-\bar{x}_{N|k}^s||_{S}^2\\
 &&J_{ER}(P_k^r,\mathbf{u}_{k})=\sum_{i=0}^{N-1}[\text{Trace}(QP_{i|k}^x+RP_{i|k}^u)]\nonumber\\
 &&\ \ \ \ \ \  \ \ \ \ \ \ \ \ \  +\text{Trace}(SP_{N|k}^x)
\end{eqnarray}
{\bf Remark 2}. {\it The first term, $J_{ET}(x_k, \mathbf{u}_k)$, in (\ref{TY23}) accounts for driving the system towards the predicted steady-state, which is related to exploitation (detailed in Section 3). In contrast, the second term, $J_{ER}(P_k^r, \mathbf{u}_k)$, in (\ref{TY23}) evaluates the effectiveness of information gathering by reducing the variance of the predicted estimation through the control sequence $\mathbf{u}_k$, which is the result of exploration.}

\subsection{Optimization problem}
Since the stable point calculated from the reference signal $\bar{r}_{i|k}$ does not necessarily belong to the system constraint set, a reference point $\bar{r}_{i|k}^s$ can be artificially introduced, as outlined in the EO-MPC tracking optimal problem. Based on the AL-MPC approach, the auto-optimal MPC problem is formulated as the following constrained optimization problem
\begin{subequations}\label{Ty24}
\begin{align}
&\!\!\!\!\!\!\!\!\!\!\!\!\!\!\!\!\!\!\!\!\!\!\!\!  \min_{\alpha_k, {\bf u}_k,\bar{r}_{i|k}^s,\bar{x}_{i|k}^s,\bar{u}_{i|k}^s}\ J_{DC}(x_k, {\bf u}_k)+\sum_{i=1}^N||\bar{r}_{i|k}^s-\bar{r}_{i|k}||_D^2 \nonumber\\
\text{s.t.}~& x_{i+1|k}=Ax_{i|k}+Bu_{i|k},\\
&\left[\begin{array}{cc}
	A& B\\
	C &0\
\end{array}\right]\left[\begin{array}{c}
	\bar{x}_{i|k}^s\\
	\bar{u}_{i|k}^s\
\end{array}\right]=\left[\begin{array}{c}
	\bar{x}_{i+1|k}^s\\
	\bar{r}_{i|k}^s\
\end{array}\right]\\
&\left[\begin{array}{cc}
	A& B\\
	C &0\
\end{array}\right]\left[\begin{array}{c}
	\bar{x}_{N|k}^s\\
	\bar{u}_{N|k}^s\
\end{array}\right]=\left[\begin{array}{c}
	\bar{x}_{N|k}^s\\
	\bar{r}_{N|k}^s\
\end{array}\right]\\
&(\bar{x}_{i|k}^s,\bar{u}_{i|k}^s)\in \mathbb{Z}, (x_{i|k},u_{i|k})\in \mathbb{Z}, x_{N|k}\in \alpha_k\mathbb{X}_f(\bar{x}_{N|k}^s).
\end{align}
\end{subequations}
The design of the terminal set and the maintenance of PE in the AL-MPC optimization problem follow the strategies outlined in Section 3. The cost functional consists of two main terms, see remark 2 for an explanation of the first term.  The second term, the offset cost functional, solely penalizes the deviation of the artificial steady-output \(\bar{r}_{i|k}^s\) from the target output \(\bar{r}_{i|k}\). Its primary purpose is to ensure that the system converges towards the desired operating condition without contributing to the exploration process. By minimizing the difference between the artificial steady-output and the target output, the system is guided towards the optimal state, thereby maintaining efficient control performance and ensuring convergence towards the predefined goals.

\subsection{Recursive feasibility and convergence of AL-MPC}
The estimated steady-state may vary at each sampling time, which presents challenges to both feasibility and convergence analysis. To tackle this issue, a two-step approach is employed. First, we analyze the recursive feasibility and convergence in the baseline case (see Section 3), where the steady-state estimate is subject to changes. Then, we examine the influence of randomness in the estimation process and its impact on the overall performance of the system.

The recursive feasibility and convergence analysis for the optimization problem in (\ref{Ty24}) are presented in Theorem 3 and Theorem 4, as outlined below.

\begin{theorem}
 If there is a feasible solution for the optimization problem (\ref{Ty24}) at time $k$, then it is recursively feasible.
\end{theorem}
{\sc Proof} At time instant $k$, assume the optimal solution  given by the control sequence $\mathbf{u}_{k}^*=[u_{0|k}^*,u_{1|k}^*,\cdots,u_{N-1|k}^*]$ with the predicted state trajectory  $\mathbf{x}_{k}^*=[x_{0|k},x_{1|k}^*,x_{2|k}^*,$ $\cdots,x_{N-1|k}^*,x_{N|k}^*]$ and $\alpha_k^*$, $\bar{r}_{i|k}^{s*}$, $\bar{x}_{i|k}^{s*}$, $\bar{u}_{i|k}^{s*}$ satisfying $x_{N|k}^*-\bar{x}_{N|k}^{s*}\in \alpha_k\mathbb{X}_f$, i.e., $x_{N|k}^*\in \alpha_k\mathbb{X}_f(\bar{x}_{N|k}^{s*}) \subseteq\mathcal{X}$. At the next time step $k+1$, the sub-optimal control  sequence and the corresponding state trajectory are denoted by $\mathbf{u}_{k+1}=[u_{1|k}^*,u_{2|k}^*,\cdots,u_{N-1|k}^*, u_{N}=K(x_{N|k}^*-\bar{x}_{N|k}^{s*})+\bar{u}_{N|k}^{s*}]$ and $[x_{1|k}^*,x_{2|k}^*,\cdots,x_{N|k}^*,$ $x_{N+1|k}]$ with $x_{N+1|k}=x_{N|k+1}$  satisfying $x_{N+1|k}-\bar{x}_{N|k}^{s*}=(A+BK)(x_{N|k}^*-\bar{x}_{N|k}^{s*})\in \alpha_k\mathbb{X}_f$ .  The state $x_{N|k+1}\in \alpha_k\mathbb{X}_f(\bar{x}_{N|k}^{s*})\subseteq\mathbb{X}$ is trivially satisfied.

\begin{theorem} \label{convergence} Under Assumption 1-4, the closed loop system
(2) under the proposed AL-MPC satisfies the  conditions $\lim_{k\rightarrow \infty} y_{k}=r^*$, implying that the system output converges to the optimal operating condition.
\end{theorem}
{\sc Proof:}  Robust set-based parameter estimation indicates that $\bar{r}_{k}^s$ asymptotically converges to the true value $f(\theta^*)$, $J_{ER}=0$. Similar to the proof of convergence presented in Section 3, $J_{ET}$ in (53)  and $J_k$ in (40)  have the same structure and are both continuous function, it follows from Lemma 3 that there exists a $\mathcal{K}_\infty$ function $\alpha(\cdot)$ such that
\begin{equation}
~~~~~~~J_{ET}^*(k)-J^*(k)\leq \alpha\left( |\bar{x}_k^s - x^s| + |\bar{r}_k^s - r^*| \right),
\end{equation}
for all $x_k\in \mathbb{X}, \hat{x}_k^s\in \mathbb{X}$ and $x^s\in \mathbb{X}$.
Therefore,  for $k\rightarrow\infty$, $\alpha(0)=0$, we have
$\lim_{k\rightarrow \infty} \bar{x}_k^s =x^s, i.e., \lim_{k\rightarrow \infty} y_{k}=r^*.$
\begin{figure*}[htbp]
    \centering
    \begin{minipage}{0.8\textwidth}
        \centering
        \includegraphics[width=\textwidth]{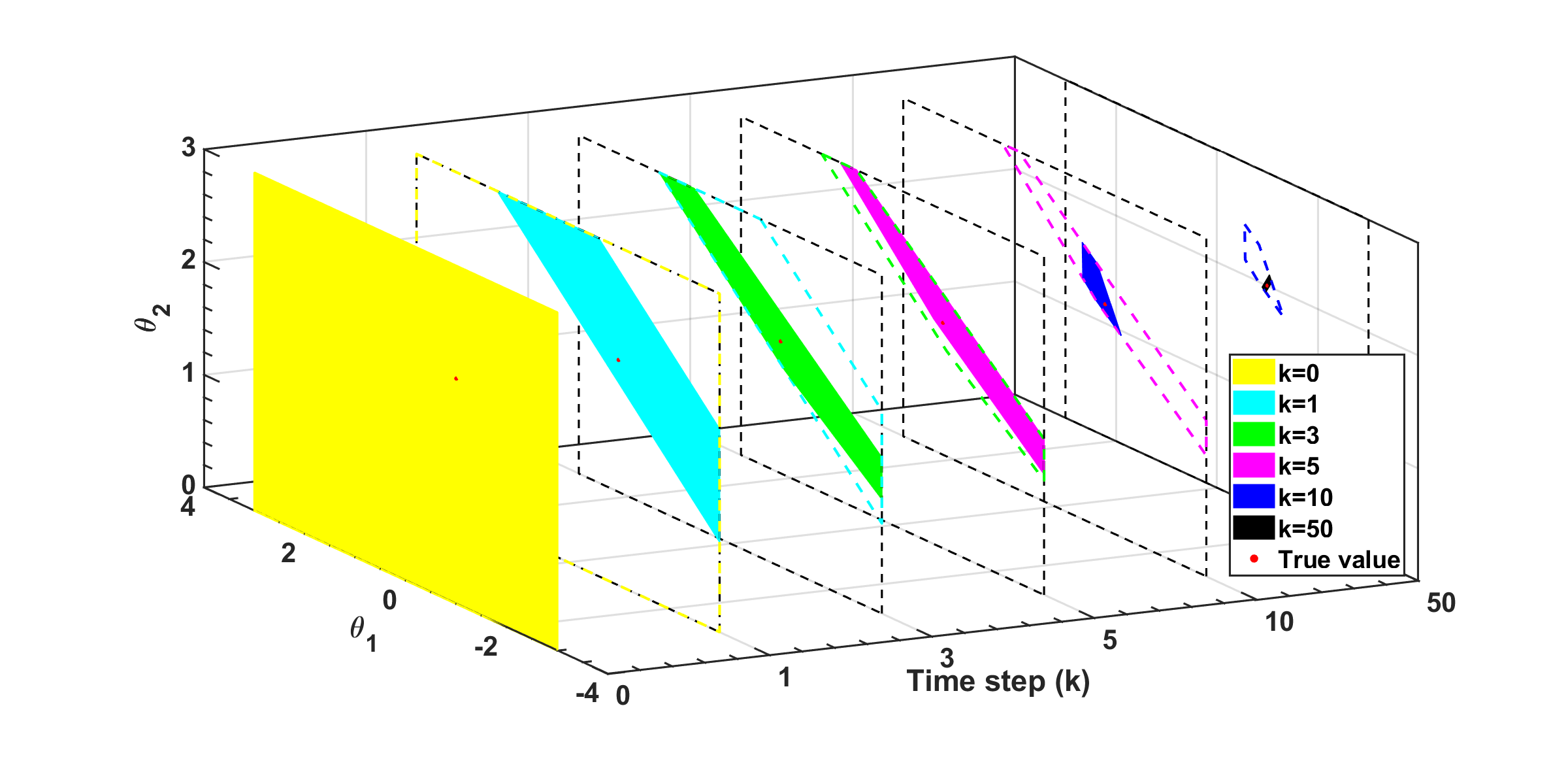}
        {(a) EO-MPC}
    \end{minipage}
    \begin{minipage}{0.8\textwidth}
        \centering
        \includegraphics[width=\textwidth]{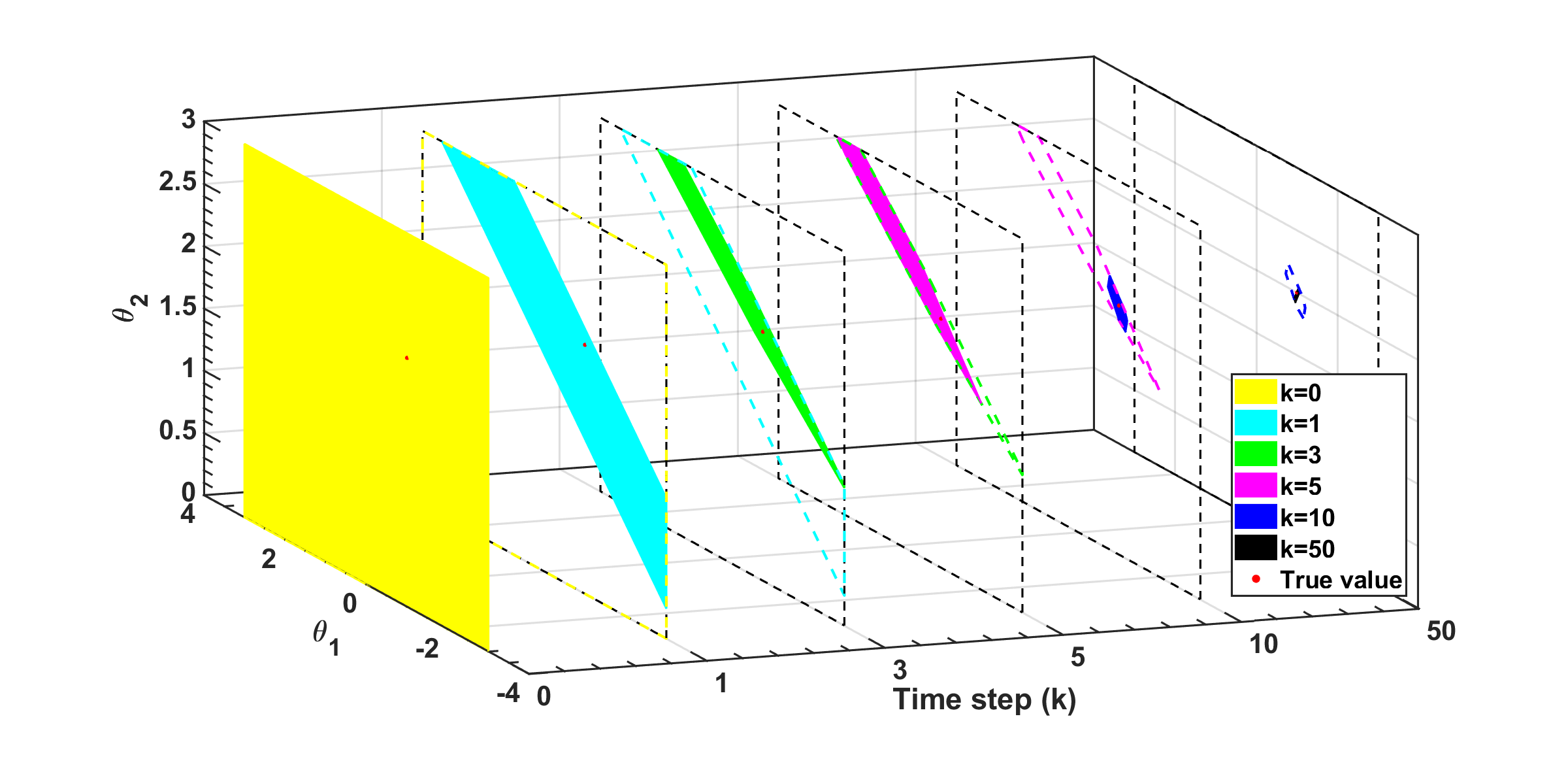}
        {(b) AL-MPC}
    \end{minipage}
\caption{Example 1: estimated parameter set at time steps $k \in \{0,1, 3, 5, 10, 50\}$ for different algorithms.}\label{TT21}
\end{figure*}
\begin{figure}
\centering
\includegraphics*[width=7cm]{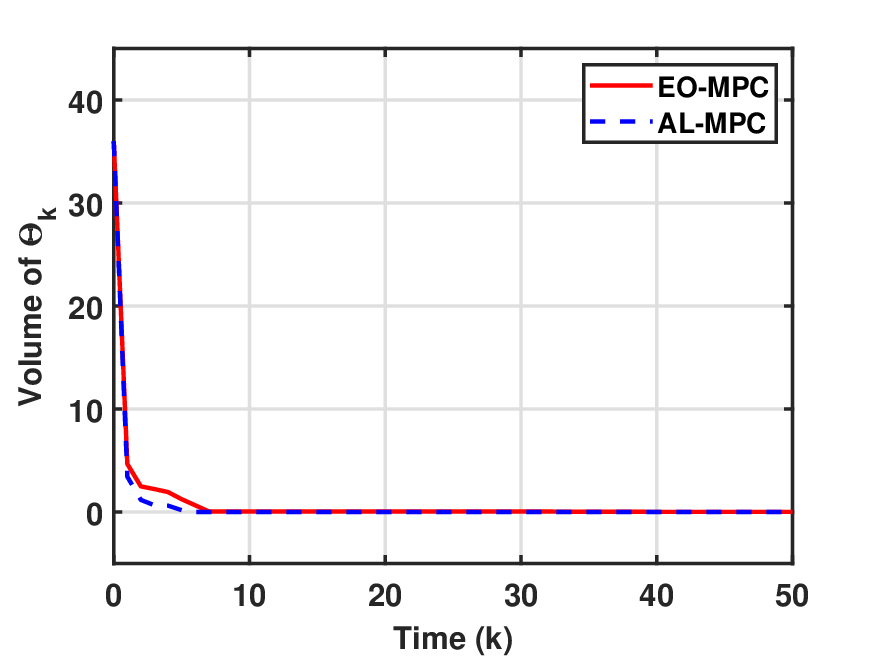}

(a) Volumes of estimated parameter sets

\centering
\includegraphics*[width=7cm]{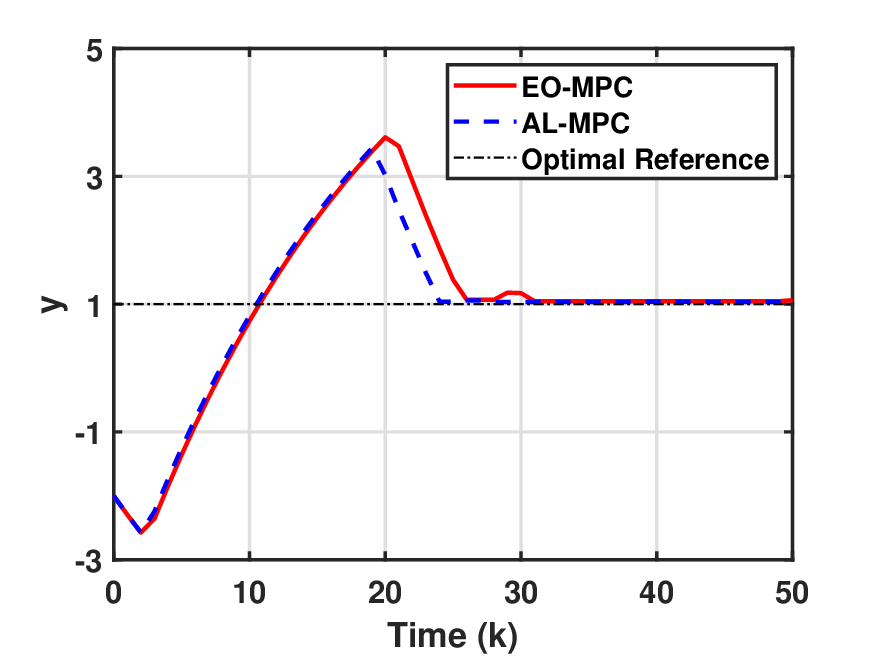}

(b) System output trajectories

\centering
\includegraphics*[width=7cm]{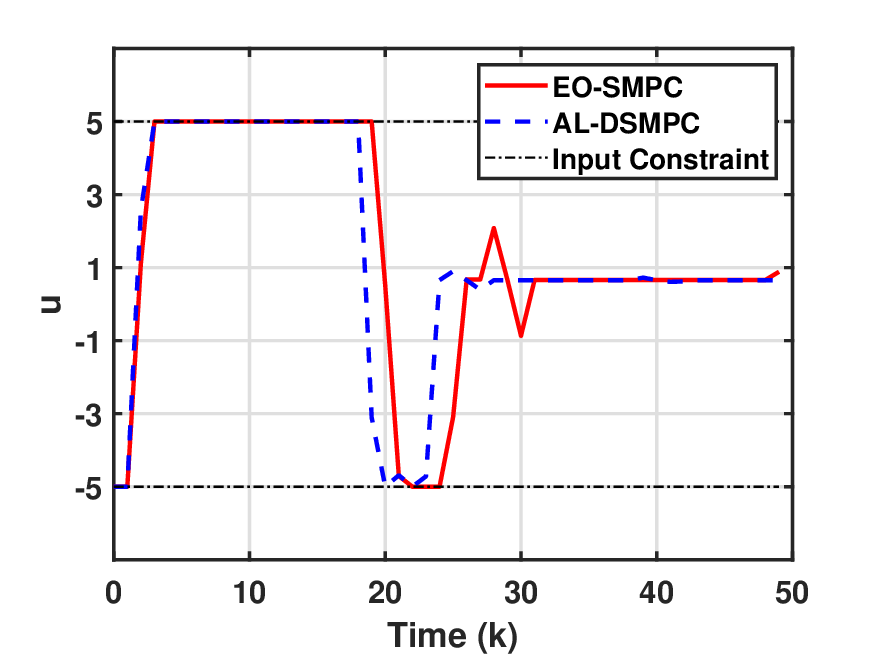}

(c) System input trajectories

\caption{Example 1: simulation results for the numerical example.}\label{TT32}
\end{figure}
\section{Examples}
\subsection{Example 1 - A numerical one}
A numerical example is considered, the matrix parameters of dynamic system (\ref{eq1a})-(\ref{eq1b}) are
$$A=\left[\begin{array}{cc}
	1.1 & 2\\
	0 & 0.95
\end{array}\right], B=\left[\begin{array}{c}
	0 \\
	0.079
\end{array}\right], C=\left[\begin{array}{cc}
	0 & 1
\end{array}\right],$$
with state and control constraints $\mathbb{X}=\{x \in \mathbb{R}^2 \mid |x| \leq 25\}$ and $\mathbb{U}=\{u \in \mathbb{R} \mid |u| \leq 5\}$. The measurement model (6) is given by
$$\phi^T(y_k)\theta^*=\theta_1^*+\theta_2^*y_k-y_k^2$$
 with environmental parameter $\theta^* = [-1 \ 2]^T$, when \( y_k \) reaches \( r^* = f(\theta^*) = \left[ 0 \ \frac{1}{2} \right] \theta^* \), \( \phi^T(y_k)\theta^* \) is maximized. The noise set is $\mathbb{V}=\{v_k \in \mathbb{R} \mid |v_k| \leq 1\}$, and the initial parameter set is $\Theta_0 = \{\hat{\theta} \in \mathbb{R}^2 \mid ||\hat{\theta}|| \leq 3\}$.

Figs. \ref{TT21} and \ref{TT32} provide a comprehensive comparison of the EO-MPC and AL-MPC algorithms, highlighting their performance in parameter estimation, system output, and input trajectories. Fig. \ref{TT21} illustrates the evolution of the estimated parameter sets at various sampling times, showing a gradual reduction in the size of the uncertainty sets as the parameter falsification process progresses. This reduction reflects the overall effectiveness of both algorithms, Fig. \ref{TT32} (a) offers a more detailed comparison by depicting the volumes of the estimated parameter sets. The AL-MPC algorithm demonstrates a significantly faster reduction in uncertainty compared to EO-MPC, driven by its active learning strategy that efficiently explores the parameter space. This leads to quicker identification and exclusion of invalid regions, enhancing the precision and speed of parameter estimation.

Figs. \ref{TT32} (b) and \ref{TT32} (c) further illustrate the system's output and input trajectories, respectively. In particular, Fig. \ref{TT32} (b) shows that the dual control strategy employed by AL-MPC successfully guides the system output to the optimal operational condition \( r^* = 1 \), minimizing tracking errors through more accurate and rapid parameter estimation. The enhanced tracking performance of AL-MPC is evident, as the algorithm converges faster and exhibits greater resilience to noise, further reinforcing its advantages in dynamic and uncertain environments.

Together, these results underscore the superiority of AL-MPC in achieving higher accuracy and operational efficiency with fewer iterations, demonstrating its robustness in addressing complex parameter estimation and control tasks.

\subsection{Example 2 - Photovoltaic Maximum Power Point Tracking}
To demonstrate the broader applicability of the proposed EO-MPC and AL-MPC strategies, we applied them to Maximum Power Point Tracking (MPPT) in photovoltaic (PV) systems, where maximizing power extraction is essential \cite{42,43}. Despite advancements, PV efficiency remains limited by environmental uncertainties such as temperature fluctuations and varying solar irradiance. We implemented EO-MPC and AL-MPC for MPPT and compared their performance with two conventional methods: perturbation and observation (PO) \cite{jung2005improved} and Q-table-based reinforcement learning (RL) \cite{kofinas2017reinforcement}.

The MPPT model was established as $P = \phi^T(V)$, where \(\phi^T(V)\) represents the polynomial regressor \([1, V, V^2, \dots, V^n]\), and \(\theta \in \mathbb{R}^{n+1}\) are the polynomial coefficients. In this application, $n=10$, to maximize power, the parameters \(\theta\) are estimated, and the voltage \(V\) is regulated by $V_{k+1} = V_k + u_k$.

\begin{figure}
\centering
\includegraphics*[width=7cm]{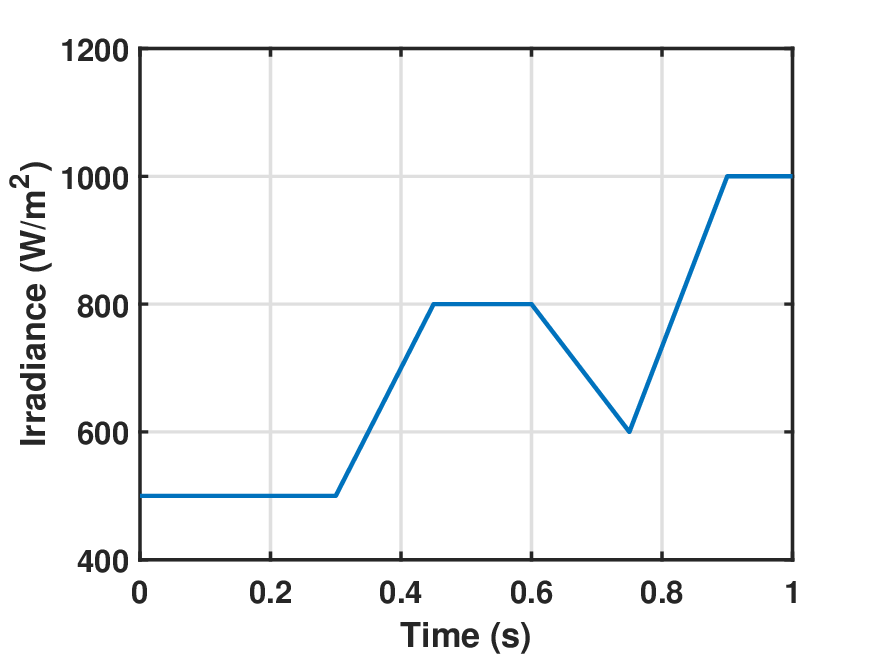}

(a) Time-varying solar irradiance profile

\centering
\includegraphics*[width=7cm]{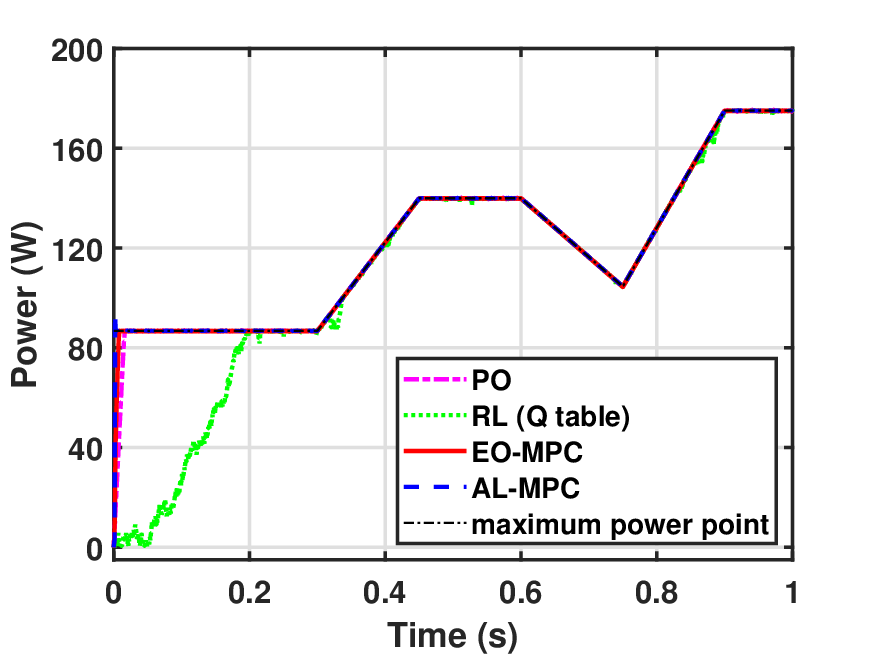}

(b) Global power profile for different algorithms

\centering
\includegraphics*[width=7cm]{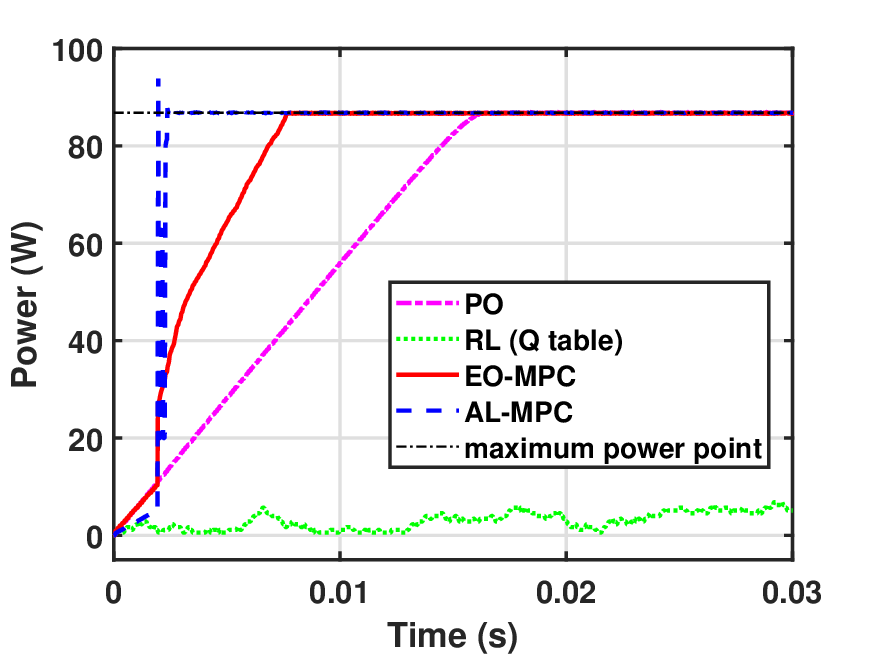}

(c) Localized power profile for different algorithms

\caption{Example 2: simulation results for  MPPT.}\label{T11}
\end{figure}

Simulation results for MPPT are shown in Fig. \ref{T11}, Fig. \ref{T11} (a) presents a time-varying solar irradiance profile.
In Fig. \ref{T11} (b)-(c), AL-MPC demonstrates superior tracking accuracy in comparison to the other methods, highlighting its effectiveness in dynamic environments. The PO method, while commonly used, induces significant voltage and current fluctuations in steady-state operations.  In contrast, RL(Q-tables) has a slow convergence rate due to the extensive data collection required for accurate state-action value estimates. This approach demands a prolonged interaction with the environment, delaying convergence to an optimal policy. While the EO-MPC is effective, AL-MPC provides a more robust solution for MPPT in PV systems, particularly under fluctuating environmental conditions. Its ability to actively explore and adapt to changes results in superior performance, making it better suited for environments where uncertainty is a key challenge.

\subsection{Example 3 - Phototropic control of nano drones}
This subsection explores the application of the proposed control methods to phototropic control in nano drones \cite{Duisterhof}. The objective is to guide a nano drone towards a light source by using a custom light sensor and control strategies such as RL(Q-table), EO-MPC, and AL-MPC.

The simulation environment was designed to mimic real-world conditions with varying light intensities and dynamic light source positions. The drone's task was to navigate towards the light source, adjusting its trajectory based on sensor readings. The light source was modeled using an inverse-square law for light intensity
\begin{equation}\label{light_model}
I(x,y) = \frac{P}{4 \pi \left((x - x_s)^2 + (y - y_s)^2 + d^2\right)},
\end{equation}
where \((x, y)\) represents the position of the drone, \((x_s, y_s)\) denotes the position of the light source, \(d\) is a small constant added to prevent singularities when the drone is close to the light source, and \(P\) is the power of the light source.

The goal of the phototropic control strategy is to maximize the light intensity $I(x,y)$ perceived by the nano drone, effectively guiding it towards the light source. The drone's position is updated by $v_{k+1} = v_k + u_k$.

\begin{figure}
\centering
\includegraphics*[width=7cm]{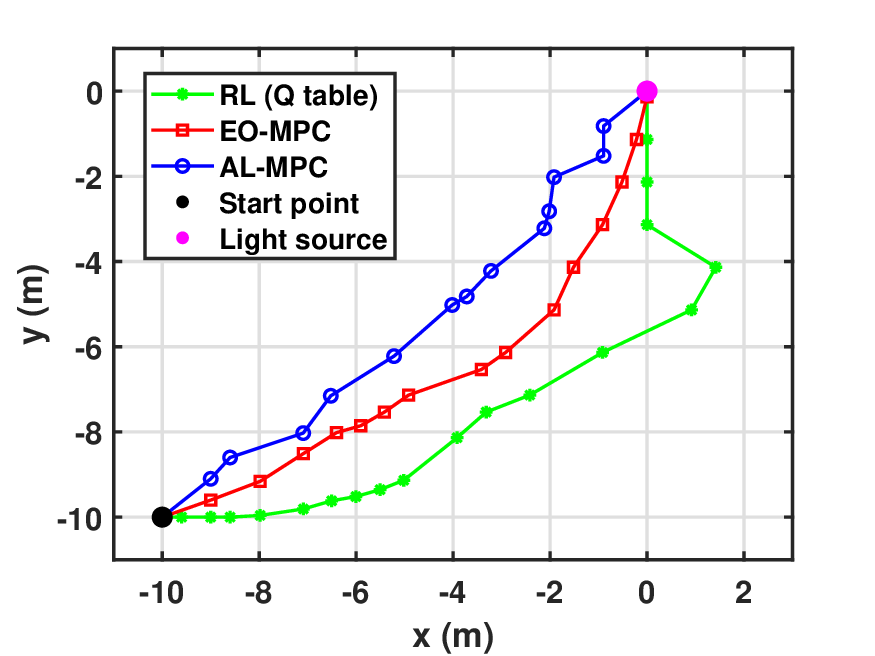}

(a) Position trajectories  for different algorithms

\includegraphics*[width=7cm]{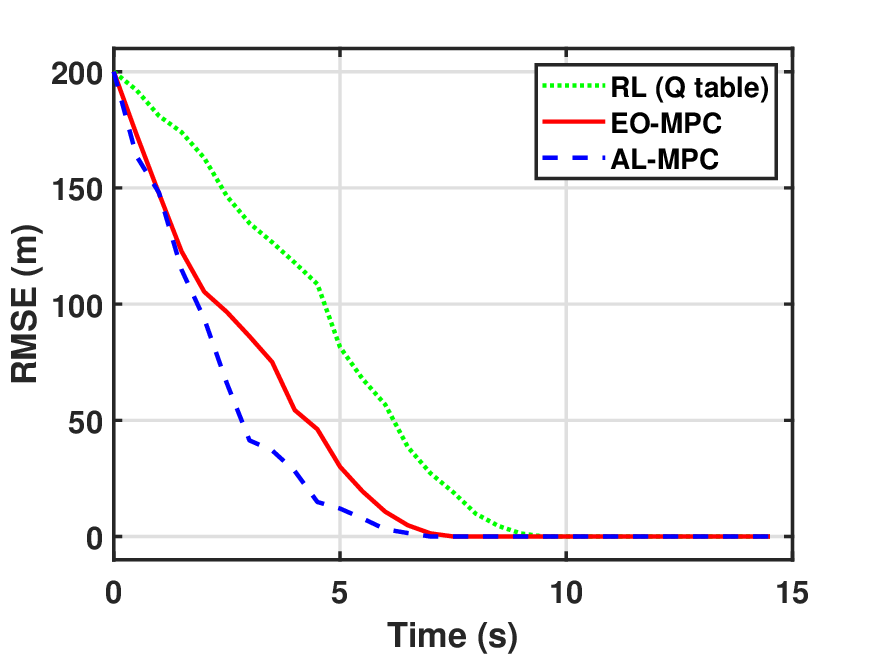}

(b) The average root-mean-square error (RMSE) for different algorithms

\caption{Example 3: simulation results for phototropic control.}\label{P1}
\end{figure}

The simulation results in Fig. \ref{P1} highlight the performance differences between RL (Q-table), EO-MPC, and AL-MPC methods for phototropic control in nano drones. Fig. \ref{P1}(a) shows the learning trajectory of the Q-table method, revealing its slow convergence due to the need to exhaustively populate the state-action space. This method's heavy reliance on direct experience with each state-action pair leads to prolonged learning phases and limits adaptability in dynamic environments. In contrast, EO-MPC  achieves faster convergence by leveraging real-time sensor data to predict future states and optimize actions immediately. Unlike the Q-table method, EO-MPC focuses on exploitation of known information, resulting in quicker attainment of the light source. However, its lack of exploratory behavior reduces adaptability in highly variable conditions.

AL-MPC balances exploration and exploitation, leading to both rapid convergence and superior adaptability, as reflected in Fig. \ref{P1} (b). This method dynamically refines its estimates while navigating changing conditions, efficiently adapting to light source shifts without extensive data collection. The path efficiency analysis shows that while the Q-table method exhibits inefficient, meandering paths and EO-MPC improves but remains suboptimal, AL-MPC consistently produces the most direct and efficient trajectories.

\section{Conclusion}
In this paper, we introduced an auto-optimal MPC in uncertain environments: AL-MPC. AL-MPC builds upon the foundation of the proposed EO-MPC. By utilizing robust set-based parameter estimation, AL-MPC effectively manages real-time uncertainty, ensuring both robust and efficient control. Through extensive simulations and practical applications, including MPPT in photovoltaic systems and phototropic control in nano drones, AL-MPC demonstrated superior performance in tracking accuracy, convergence speed, and adaptability compared to traditional methods.

\end{document}